%Paper: hep-th/9502107
%From: Andy Strominger <andy@denali.physics.ucsb.edu>
%Date: Thu, 16 Feb 95 15:03:34 PST

\input phyzzx
\Pubnum={\vbox{\hbox{NSF-ITP-95-07}\hbox{hep-th/9502107}}}
\date={February 1995}
\pubtype={}
\def\a{\alpha}
\def\b{\beta}
\def\g{\gamma}
\def\d{\delta}
\def\e{\epsilon}
\def\ebar{\e_-}

\def\et{\eta}

\def\l{\lambda}
\def\m{\mu}
\def\n{\nu}

\def\r{\rho}

\def\s{\sigma}

\def\p{\phi}

\def\c{\chi}
\def\ps{\psi}

\def\D{\Delta}

\def\S{\Sigma}

\def\o{\over}
\def\ri{\rightarrow}

\def\h{{1\over 2}}
\def\np{{\it Nucl. Phys. }}
\def\pl{{\it Phys. Lett. }}
\def\cmp{{\it Comm. Math. Phys. }}

\def\pr{{\it Phys. Rev. }}
\def\rmp{{\it Rev. Mod. Phys.}}

\def\b{\beta}
\def\g{\gamma}

\def\a{\alpha}

\def\D{\Delta}
\def\e{\varepsilon}
\def\no{\noindent}

\def\IR{\relax{\rm I\kern-.18em R}}
\font\cmss=cmss10 \font\cmsss=cmss10 at 7pt
\def\IZ{\relax\ifmmode\mathchoice
{\hbox{\cmss Z\kern-.4em Z}}{\hbox{\cmss Z\kern-.4em Z}}
{\lower.9pt\hbox{\cmsss Z\kern-.4em Z}}
{\lower1.2pt\hbox{\cmsss Z\kern-.4em Z}}\else{\cmss Z\kern-.4em Z}\fi}
\def\IN{\relax{\rm I\kern-.18em N}}
\def\Dsl{{\cal D}\!\!\!\!/}
\def\dsl{\partial \!\!\!/}

\def\zbar{\bar z}
\def\pa{\partial}
\def\Dt{{\cal D}}

\titlepage
\title{THREE-DIMENSIONAL SUPERGRAVITY AND THE COSMOLOGICAL CONSTANT}
\vskip 0.25cm
\author{Katrin Becker}
\address{Institute for Theoretical Physics\break University of California
\break Santa Barbara, CA 93106-4030}
\author{Melanie Becker \break and \break Andrew Strominger}
\address{Department of Physics\break University of California \break Santa
Barbara, CA 93206-9530}

\abstract{Witten has argued that in $2+1$ dimensions local supersymmetry
can ensure the vanishing of the cosmological constant without requiring the
equality of bose and fermi masses. We find that this mechanism is implemented
in a novel fashion in the (2+1)-dimensional
supersymmetric abelian Higgs model coupled to supergravity.
The vortex solitons are annihilated by half of the supersymmetry
transformations. The covariantly constant spinors required to define these
supersymmetries
exist by virtue of a surprising cancellation between the Aharonov-Bohm phase
and the phase associated with the holonomy of the spin connection.
However the other half of the supersymmetry transformations, whose
actions ordinarily
generate the soliton supermultiplet, are not well-defined and bose-fermi
degeneracy is consequently absent in the soliton spectrum.}

\endpage
The cosmological constant problem is surely one of the most vexing problems in
all of physics\foot{An excellent review of  the cosmological constant problem
can be found in ref. \REF\weinberg{S.~Weinberg, {\rmp} {\bf 1} (1989) 61.}
[\weinberg]}. Supersymmetry has something to say about the problem, but it is
not clear if it makes matters better or worse. In some theories ({\it e.g.}
string theory) the cosmological constant can naturally vanish before
supersymmetry breaking. However after supersymmetry breaking it is typically
non-zero, and cannot be made to vanish even by fine-tuning bare parameters.

Some time ago \REF\witten{E.~Witten, ``Is Supersymmetry Really Broken?'', talk
given at the {\it ``Santa Fe Institute Meeting on String Theory''}, 1985,
preprint IASSNS-HEP-94-72, hep-th/9409111.}[\witten] Witten made a striking
observation concerning the relationship between local supersymmetry and the
cosmological constant in $2+1$ dimensions. The vacuum can have exactly zero
cosmological constant because of local supersymmetry, yet the excited states
may not be
in degenerate bose-fermi pairs. This is because supercharges -- whose existence
ordinarily implies bose-fermi degeneracies -- are ill-defined in the conical
geometries arising in non-zero energy states in $2+1$ dimensions
\REF\deser{S.~Deser, R.~Jackiw and G.~`t Hooft, {\it Ann. Phys.} {\bf 152}
(1984) 220.}[\deser]. Thus in $2+1$ we can have our cake and eat it, too. It
would certainly be wonderful if a $3+1$ dimensional theory in which
supersymmetry implied zero cosmological constant but not the unwanted
degenerate bose-fermi pairs could be found! Unfortunately there have been no
suggestions of how to
implement this idea in $3+1$.

In this paper we investigate this mechanism as it applies to the
solitons of the $N=2$
supersymmetric abelian Higgs model coupled to supergravity
in 2+1 dimensions. Before coupling to supergravity
this theory has a supermultiplet of Nielsen-Olesen vortex solitons
\REF\nielsen{H.~B.~Nielsen and P.~Olesen, \np {\bf B61} (1973) 45.} [\nielsen]
of mass $M$.
The solitons are
annihilated by half of the supersymmetry
transformations. The action of the other, broken, half generates
fermionic Nambu-Goldstone zero modes. Quantization of these zero modes leads to
a supermultiplet of degenerate bosonic and fermionic soliton states.
We shall find that the unbroken supersymmetries ingeniously
survive the coupling to supergravity, despite the existence of a conical
geometry. This is possible because in the locally supersymmetric theory the
supersymmetry transformation parameter becomes charged. The geometric phase
associated to the conical geometry
is then cancelled by an Aharonov-Bohm phase.
On the other hand the phases add rather than cancel for the would-be broken
supersymmetry generators. There are accordingly no normalizable fermion zero
modes, and the bose-fermi degeneracy is split.

Curiously this model contains fermionic particles with
potentially fractional charges $v^2g/2M_p$, where $g$ is
the Higgs charge, the constant $v$ is the magnitude of the Higgs vacuum
expectation value, $v^2$ is the coefficient of the
Fayet-Iliopoulos $D$-term and $M_p$ is the Planck Mass. If topologically
non-trivial
gauge connections\foot{{\it i.e.} connections for which the integral of the
field strength $F$ over a closed two surface is non-zero.} are allowed, and
$g$ is an integer multiple $p$ of the fundamental electric charge,
then
the coupling to supergravity is only consistent if $v$ is quantized according
to $pv^2=2qM_p$, where $q$ is an integer. A similar quantization condition
\foot{This quantization condition is related to the non-renormalization
theorems for the Fayet-Iliopoulos $D$-term discussed in the context of string
theory
\REF\string{M.~Dine, N.~Seiberg and E.~Witten, \np {\bf B289} (1987) 589;
J.~Atick, L.~Dixon and A.~Sen, \np {\bf B292} (1987) 109;
M.~Dine, I.~Ichinose and N.~Seiberg, \np {\bf B293} (1987) 253.}[\string].
The situation encountered in string theory is somewhat different in that
a non-linear U(1) transformation law for the axion
(associated with anomaly cancellation) cancels the shift in
U(1) fermion charges.} applies to the four-dimensional Abelian Higgs model
coupled to supergravity but does not appear to have been previously noticed.

We first review the solitons of the $(2+1)$-dimensional abelian Higgs model
with $N=2$ global supersymmetry. The Lagrangian is
\REF\vecchia{P.~di~Vecchia and S.~Ferrara, \np {\bf B130} (1977) 93.}
\REF\lee{R.~Jackiw and C.~Rebbi, \pr {\bf D13} (1976) 3398; R.~Jackiw, K.~Lee
and E.~Weinberg, \pr {\bf D42} (1990) 3488; C.~Lee, K.~Lee and H.~Min, \pl {\bf
B252} (1990) 79.}
\REF\llm{B.~H.~Lee, C.~Lee and H.~Min, \pr {\bf D45} (1992) 4588.}
[\vecchia,\lee,\llm]

$$
\eqalign{
{\cal L}=&-{1\o 4}F_{\m\n}F^{\m\n}-{\cal D}_\m\p{\cal D}^\m\p^*-\h \partial_\m
N \partial^\m N-g^2 N^2 \p\p^*-{g^2 \o 2} D^2(\p)\cr
&-i{\bar \chi}\Dsl \chi -i {\bar \l}\dsl \l+i \sqrt{2}g\left({\bar \chi} \l \p
-{\bar \l} \chi \p^* \right)-gN{\bar \c} \c,\cr}
\eqn\ai
$$
where
$$
D(\p)=\p^* \p -v^2.
\eqn\aii
$$
Here $g$ is the gauge coupling (with dimensions of ${\rm (mass)}^{1/2}$), $N$
is a neutral real scalar, $\p$ is a complex charged scalar, $\chi\;(\l)$ is a
complex charged (neutral) two-component spinor, ${\cal
D}_\mu=\partial_\mu-igA_\mu$ is the covariant derivative when acting on $\p$
and ${\bar \chi}=\chi^{\dagger}\g^0$, {\it etc.}. The metric tensor $\eta_{\mu
\nu}$ has the signature $(-,+,+)$. The $\g$ matrices
can be represented by $\g^0=\s^3$, $\g^1=i\s^2$ and $\g^2=i\s^1$ and they
satisfy the relation $\g^\mu \g^\nu=-\eta^{\mu \nu} -i \e^{\mu \nu \l} \g_\l$.

This theory is invariant under $N=2$ supersymmetry transformations:
$$
\eqalign{
&\d_{\e}\c=i\sqrt{2}\g^\m{\cal D}_\mu\p\e -\sqrt{2}gN\p\e ,\cr
& \d_\e\l=F_{\m\n}\g^{\m\n}\e-igD(\p)\e-\pa_\m N\g^\m \e,\cr
&\d_\e A_\m=i({\bar \e}\g_\m\l-{\bar \l}\g_\m\e),\cr
&\d_\e\p=\sqrt{2}{\bar \e}\chi,\cr
&\d_\e N=i({\bar \l}\e-{\bar  \e}\l).\cr}
\eqn\aiii
$$
Here the parameter $\e$ is a complex anticommuting spinor, $\g^{\mu\nu}={1\o
4}[\g^\mu,\g^\nu]$.

The soliton is given by the static, vortex field configuration obeying the
first order differential equations
$$
F\equiv\e^{z\zbar}F_{z\zbar}=gD(\p)\qquad{\rm and} \qquad  {\cal D}_z\p=0,
\eqn\aiv
$$
where $(z,\zbar)$ are complex spatial coordinates, $\e^{z\zbar}=-\e^{\zbar
z}=-i\et^{z\zbar}=-2i$ in flat space and
%. We consider field configurations satisfying:
$$
F_{0z}=F_{0\zbar}=N=0.
\eqn\aaaiv
$$
Antivortices with $F=-gD(\p)$ and $\Dt_z\p^*=0$ also exist.

The solutions of the first order differential equations {\aiv} are labeled by
the magnetic flux
$$
\int F d^2 z =-{4\pi\o g} n\qquad {\rm with}\quad n \in \IZ_+,
\eqn\aaiv
$$
while antivortices satisfy equation {\aaiv} with $n\in \IZ_-$. In the following
we take $n$ positive.

The soliton solutions are known to exist but they cannot be found analytically
by solving equation {\aiv}. However what is important for our considerations is
that
all local gauge invariant quantities fall to zero exponentially outside of a
core region of characteristic size $1/v^2$, because there are no massless
propagating
fields.

The solution {\aiv} breaks half of the supersymmetries of the theory. To see
this decompose $\e$ into spinors $\e_+$ and $\ebar $ of definite spatial
chiralities:
$$
\eqalign{
&\g^{\zbar}\e_+ =0,\cr
&\g^z \e_-=0.\cr}
\eqn\av
$$
Then it is easy to see that the transformations generated by $\e_+$ are
unbroken
$$
\d_{\e_+}\c=\d_{\e_+}\l=0,
\eqn\avi
$$
The transformations generated by $\ebar $ are broken
$$
\eqalign{
&\d_{\e_-}\c=i\sqrt{2}\g^{\zbar}\left({\cal D}_{\zbar}\p\right)\e_-
\neq 0, \cr
&\d_{\e_-} \l =-2igD(\p) \e_- \neq 0, \cr}
\eqn\avii
$$
except when we are in the vacuum where $D(\p)=0$, {\it i.e.} $|\p|=v$.
By virtue of Goldstone's theorem we expect a massless excitation for every
broken symmetry. Indeed {\avii} are the Nambu-Goldstone zero modes in the
soliton background. They are normalizable because of the exponential falloff
of $D(\p)$ and ${\cal D}_{\zbar}\p$.

The solitonic spectrum is obtained by quantizing
these zero modes. The operator $b_0$ which creates a fermion zero mode obeys:
$$
\eqalign{
& \{ b_0^*, b_0\} =1,  \cr
& \{ b_0, b_0\} =\{ b_0^*, b_0^*\}=0\cr}
\eqn\aviii
$$
and of course carries no energy. The soliton groundstate for $n=1$
is then a representation of {\aviii} corresponding to a massive $(0,1/2)$
supermultiplet.

For the model with global supersymmetry, we can conclude that the cosmological
constant vanishes because all the supersymmetries are unbroken in the vacuum
\REF\zumino{B.~Zumino, \np {\bf B89} (1975) 535.} [\zumino]. However, there is
a phenomenologically undesirable degeneracy between bosons and fermions. This
situation will change for the model with local supersymmetry.

The locally supersymmetric action in $2+1$ dimensions does not seem to appear
explicitly in the literature but may be obtained by reduction of the
four-dimensional $N=1$ abelian Higgs model coupled to supergravity
\REF\bagger{A.~Das, M.~Fischler and M.~Ro\v cek, \pr {\bf D16} (1977) 3427;
D.~Z.~Freedman and J.~H.~Schwarz, {\pr} {\bf D15} (1977) 1007;
K.~S.~Stelle and P.~C.~West, \np {\bf B145} (1978) 175; D.~E.~Cremmer,
B.~Julia, J.~Scherk, S.~Ferrara, L.~Girardello and P.~van Nieuwenhuizen, \np
{\bf B147} (1979) 105; J.~A.~Bagger, \np {\bf B211} (1983) 302; E.~Cremmer,
S.~Ferrara, L.~Girardello and A.~van Proeyen, \np {\bf B211} (1983) 413.}
\REF\wess{See e.g. J.~Wess and J.~Bagger, ``Supersymmetry and Supergravity'',
Princeton Univ. Press, Jun 1990.}[\bagger,\wess].
The resulting Lagrangian is
$${\cal L}=
{M_p\o 2}R-{1\o 4}F_{\m\n}F^{\m\n}-\Dt_\m\p\Dt^\m\p^*
-\h \Dt_\m N\Dt^\m N
-g^2N^2\p^* \p-{g^2\o 2}D(\p)^2
+({\rm fermi}).
\eqn\ax
$$
where $\Dt_\mu$ is the covariant derivative with respect to gravity and the
U(1) gauge group. The full Lagrangian is invariant under local supersymmetry
transformations and under the U(1) gauge transformations
$$
\eqalign{
&\d_\a\p=i\a\p,\cr
&\d_\a A_\m =g^{-1}\partial_\m \a,\cr
&\d_\a\c=i\a\c+{iv^2\a \o 2 M_p}\c,\cr
&\d_\a\l= {i v^2 \a\o 2 M_p}\l,\cr
&\d_\a\ps_\m= {i v^2\a \o 2 M_p}\ps_\m.\cr}
\eqn\aaax
$$
 From this formula we observe that the gravitino is charged while the charges
of
other fermions are shifted when supergravity is coupled. These charges have
their origin in the K\"ahler invariance of the Lagrangian
\REF\baggerwitten{J.~Bagger and E.~Witten, \pl {\bf B118} (1982) 103;
J.~Bagger, ``Supersymmetric Sigma Models'', Lectures given at the ``{\it
Bonn-Nato Advanced Study Institute on Supersymmetry}'', 1984, preprint
SLAC-PUB-3461.}[\baggerwitten].
If topologically non-trivial connections are allowed\foot{The vortex solution
on its own is not incompatible with fractional charge because all of spacetime
can be covered with one patch. Problems with fractionally charged objects
arise when there are non trivial transitions between neighboring patches.}
and $g$ is an integer multiple $p$ of the minimal electric charge, charge
quantization implies $pv^2=2qM_p$ where $q$ is an integer.

The relevant part of the supersymmetry transformations becomes\footnote{*}{This
corrects a misprint in ref. [\wess].}
$$
\eqalign{
&\d_\e\c=i\sqrt{2}\g^\m\Dt_\m\p\e,\cr
&\d_\e\l=F_{\m\n}\g^{\m\n}\e-igD(\p)\e,\cr
&\d_\e\ps_\m=\Dt_\m^-\e.\cr}
\eqn\axi
$$
where we have defined the K\"ahler-covariant derivative
$$
\Dt_\n^{\pm}=\Dt_\nu\pm {J_\n\o 4M_p}
\qquad {\rm and}\qquad J_\nu=\p^*\Dt_\n \p-\p \Dt_\n \p^*.
\eqn\aaviii
$$
The supersymmetry parameter $\e$ is charged and transforms as:
$$
\d_\a\e={iv^2 \a \o 2M_p}\e.
\eqn\aaaax
$$
The bosonic equations of motion that follow from the Lagrangian are
$$
\eqalign{
M_p G_{\m\n}=T_{\m\n}\equiv&-{1\o 4}g_{\m\n} F_{\a\b}F^{\a\b}+g^{\a\b}F_{\m\a}
F_{\n\b}-g_{\m\n}\Dt_\a \phi \Dt^\a\p^*+\Dt_\m\p\Dt_\n\p^*\cr
&+\Dt_\m \p^*\Dt_\n\p-{1 \o 2}g_{\m\n}\Dt_\a N\Dt^\a N+\Dt_\n N\Dt_\m N -{g^2
\o 2}g_{\m\n}D(\p)^2,\cr
\Dt_\m\Dt^\m\p =&g^2 \p D(\p),\cr
\Dt^\m F_{\m\n} =&igJ_\n.\cr}
\eqn\aax
$$
As in flat space we will look for static solutions that satisfy
$F_{0z}=F_{0\zbar}=N=0$. For such solutions the line element can be put in the
form
$$
ds^2=-dt^2+e^{2\r}dzd{\zbar},
\eqn\aaxii
$$
and the $t$-$t$ component of Einsteins equation takes the form
$$
M_pR_{z\zbar}=-{1\o 2}e^{2\r}v^2Fg+{1 \o 2} \left(\Dt_z J_{\zbar}-\Dt_{\zbar}
J_z\right)+{1\o 4}e^{2\r} (F-gD(\p))^2+2\Dt_z\p \Dt_{\zbar}\p^*.
\eqn\axiii
$$
It is then easy to see that the matter equations of motion are solved by the
Landau-Ginzburg vortices
$$
\eqalign{
&F=gD(\p)\qquad{\rm and}\qquad\Dt_z\p=0.\cr}
\eqn\axii
$$
Using {\axii}, equation {\axiii} reduces to a linear relation for $\r$
$$
M_pR_{z\zbar}=-2M_p \pa_z\pa_{\zbar}\r=iv^2 g F_{z\zbar}+{1\o 2}\left(\Dt_z
J_{\zbar}-\Dt_{\zbar} J_z\right) .
\eqn\aaxii
$$
Asymptotically the right hand side of equation {\axiii} vanishes, and the
geometry is therefore locally flat. The metric takes the form
$$
ds^2 \simeq -dt^2 +{dz d\zbar \o (z \zbar )^{M/M_p}}\qquad {\rm as } \qquad
r^2=z\zbar\rightarrow \infty,
\eqn\axiv
$$
where
$$
M=v^2 n
\eqn\axv
$$
(with $n>0$) is proportional to the soliton mass. This corresponds to a cone
with deficit angle $2\pi M/M_p$. From equation {\axiv} we observe that if
$M>M_p$, the metric becomes singular [\deser]. For the marginal case $M=M_p$
the space is asymptotic to a cylinder.

It is easy to see that as in the global case,
$$
\d_{\e_+}\c=\d_{\e_+}\l=0,
\eqn\avi
$$
for any positive chirality spinor $\e_+$. Half of the supersymmetries are
unbroken if we can find a specific $\e_+$ with the additional property
$\d_{\e_+}\psi_\m=0$. If $\e_+$ is time-independent, $\d_{\e_+}\psi_0$ will
trivially vanish. The conditions
$$\eqalign{
\d_{\e_+}\psi_z= \Dt_z^-{\e_+}&=\pa_z{\e_+}-{J_z\o 4M_p} {\e_+}-i{v^2 g \o
2 M_p}A_z{\e_+}=0,\cr
\d_{\e_+}\psi_{\zbar} =\Dt_{\zbar}^-{\e_+}&=\pa_{\zbar}{\e_+}
-\pa_{\zbar}\r\e_+-
{J_{\zbar}\o 4M_p}{\e_+}-i{v^2 g \o 2M_p}A_{\zbar}{\e_+}=0,\cr}
\eqn\sss
$$
are differential equations for $\e_+$. Solutions will exist if the
integrability
condition
$$
\left[\Dt_z^-, \Dt_{\zbar} ^-\right]\e_+=0
\eqn\intg
$$
is satisfied. This condition is indeed equivalent to the constraint
equation {\aaxii}, and we conclude that half of the supersymmetries are
unbroken.

This result may come as a surprise since covariantly constant
spinors do not usually exist in asymptotically conical spaces:
a  phase is acquired in parallel transport about a
circle at infinity.
However in this case the phase is cancelled
by an Aharonov-Bohm phase which arises because the gravitino
has charge! To see this explicitly note that asymptotically
as $r\to \infty$
$$
\pa_z\r \to -{M\o 2M_p z}, \qquad (A_z,A_{\zbar})\to
{in\o 2g}\left({1\o z},-{1\o \zbar}\right) \qquad {\rm and}
\qquad (J_z,J_{\zbar}) \to (0,0)
\eqn\asmt
$$
to leading order in $1/r$, where we have
used {\aaiv}, {\axiv}
and the
relation
$$
\int_\S d^2 z\partial_{\zbar}f(z, \zbar)=-i\oint_{\partial \S} dzf(z, \zbar).
\eqn\stokes
$$
Using $M=nv^2$,
cancellations occur between the
connections, and the covariant constancy
conditions {\sss}
reduce to
$$
\eqalign{
\Dt_z^-{\e_+}&\to\pa_z{\e_+} +{M\o 4 M_p z}{\e_+}=0,\cr
\Dt_{\zbar}^-{\e_+}&\to\pa_{\zbar}{\e_+} +{M\o 4 M_p {\zbar}}{\e_+}=0.\cr}
\eqn\rdc
$$
The solutions obey
$$
{\e_+}\to(z{\zbar})^{-M/4M_p}\e_0\qquad {\rm and} \qquad e^{-\r}\e_+^\dagger
{\e_+}\to {\rm constant}
\eqn\ddd
$$
as expected for a parameter which generates a nontrivial global supersymmetry
transformation.

Stability of this solution follows from a Bogomol'nyi bound
\REF\bogo{E.~B.~Bogomol'nyi, {\it Yad. Fiz.} {\bf 24} (1976) 861, {\it Sov. J.
Nucl. Phys.} {\bf 4} (1976) 449.} [\bogo] relating the mass of a configuration
to the magnetic flux:
$$
M\geq v^2|n|.
\eqn\bi
$$
The above inequality is saturated if the configurations satisfy the first order
differential equations.
This bound can be derived using a variant of the methods of refs.
\REF\witt{E.~Witten, \cmp {\bf 80} (1981) 381.}
\REF\gihu{G.~W.~Gibbons and C.~M.~Hull, \pl {\bf  109B} (1982)
190.}[\witt,\gihu]. Because of the infrared divergences, it appears necessary
to work at large, but finite $r$. Define
$$
\D(r)\equiv i\oint_{{\cal C}_r} dx^\a {\bar \et} \Dt^-_\a \et.
\eqn\bii
$$
$\et$ here is an anticommuting
spinor which transforms like $\e$ and will be further constrained below.
The integration contour ${\cal C}_r$ is a curve of fixed $r$ embedded in a
spacelike slice $\S$ on which the metric asymptotically approaches the conical
form
{\axiv}. $\D(r)$ may be expressed as a volume integral over the portion $\S_r$
of $\S$ inside ${\cal C}_r$. One finds
$$\eqalign{
\D(r)=\int_{\S_r} d^2 \S_{\mu}\lbrack & i\e^{\m \a\b} \Dt^+_\a {\bar \et}
\Dt^-_\b \et-{1 \o 2}G^{\m \a} {\bar \et} \g_\a \et+{v^2 g\o 4 M_p}
\e^{\m\a\b}F_{\a\b} {\bar \et} \et \cr &-{i \o 4M_p}{\bar \et} \et\e^{\m\a\b}
{\cal D}_\a J_\b \rbrack.\cr}
\eqn\bv
$$
It is always possible to find a coordinate system in which the metric takes the
simple form {\axiv} on $\S$ (though this can not be done throughout the
entire spacetime for nonstatic geometries). With respect to such coordinates we
further impose the condition
$$
\g^{\zbar} \et =\Dt^-_{z} \et =0,
\eqn\biv
$$
so that $\et$ reduces to a single complex component which we denote $\et_+$.
$\D(r)$ then reduces to
$$
\D(r) =\int d^2 z\lbrack e^{-\r}|\Dt^-_{\zbar}\et_+|^2+e^{\r}{1\o 8M_p} |\et|^2
(F-gD(\p))^2+ e^{-\r} {1 \o M_p}|\et|^2|\Dt_z\p|^2\rbrack,
\eqn\bvi
$$
where $|\et|^2 =\et_+^\dagger\et_+$, {\it etc.} Evidently
$$
\D(r)\geq 0.
\eqn\bvii
$$
The equality holds if and only if every term vanishes, which implies $M=v^2 n$.
On the other hand the asymptotic behavior of $\et$ can be read off from {\biv}
after inserting the asymptotic form {\asmt} of the connection
$$
\et \sim r^{-v^2n/2M_p}.
\eqn\bviii
$$
It follows that as $r\ri \infty$
$$
\D(r) \sim (M-v^2n) r^{(M-v^2n)/2M_p}.
\eqn\bix
$$
Now suppose that $M$ is less than $v^2 n $. Then $\D(r)$ is negative for large
$r$. But this contradicts {\bvii}. One may also show that
$M\geq -v^2 n$ for antivortices
with $n<0$ by considering spinors which
obey $\g^z \et =\Dt^-_{\zbar} \et=0$. We conclude that the
bound {\bi} is valid, and is saturated by the stable solution
{\axiv}.

One might expect that Nambu-Goldstone zero
modes can be constructed from the broken supersymmetries.
As in the global case
$$
\eqalign{
&\d_{\e_-}\c=i\sqrt{2}\g^{\zbar}\left({\cal D}_{\zbar}\p\right)\e_-
\neq 0, \cr
&\d_{\e_-} \l =-2igD(\p) \e_- \neq 0, \cr
&\d_{\e_-} \psi_z=\pa_z{\e_-}-\pa_z\r\e_-
-{J_z\o 4M_p}{\e_-}-i{v^2 g \o  2M_p}A_z{\e_-}\neq 0,\cr
&\d_{\e_-} \psi_{\zbar}=\pa_{\zbar}{\e_-}-
{J_{\zbar}\o 4M_p} {\e_-}-i{v^2 g \o 2M_p}A_{\zbar}{\e_-}\neq 0,}
\eqn\avzii
$$
is a zero mode. However if $\e_-$ has the asymptotic
behavior {\ddd} corresponding to a physical ({\it i.e.} not pure gauge)
supersymmetry transformation, the zero mode {\avzii} is not normalizable.
The norm of this mode has an infrared divergent contribution
$$
\int d^2 z \left( \d_{\e_-} \ps_{\mu} \right)^* \d_{\ebar} \ps_{\mu}  e^{-\rho}
\sim \int d^2 z ( z \zbar )^{-1}.
\eqn\axvii
$$
Thus in the supergravity theory the would-be Nambu-Goldstone
zero mode picks up a small, but long-range gravitino term which renders it
non-normalizable, and so it does not enter in the construction of the physical
Hilbert space\foot{An exception to this occurs when $M=M_p$, for
which the space is asymptotic to a cylinder and a normalizable zero
mode of the form {\avzii} exists.}.

While there are no physical fermion zero modes for a single soliton,
there is a lowest-lying eigenmode. As $M_p$ is made large, and
$M/M_p\rightarrow 0$ gravitational effects become very weak inside the core of
the soliton. In this limit the degenerate supermultiplet should reappear, so
one expects the lowest eigenvalue to be proportional to a power of $M/M_p$.

In the vacuum all the supersymmetries are unbroken, and the cosmological
constant vanishes. Therefore, when the supersymmetry is local,
it can imply a vanishing cosmological constant without also implying
undesirable degenerate bose-fermi supermultiplets.

In [\witten ] Witten observed that the phases arising in conical geometries
disturb the usual connection between supersymmetry of the vacuum and
bose-fermi degeneracy of the excited states. In the present paper,
this has been
explicitly verified in a specific model. This model in addition exhibits
an effect equal in importance to the conical  phases: Aharonov-Bohm
phases. Perhaps this observation might be useful for generalizing
Witten's mechanism to $3+1$ dimensions.

\vskip 1cm
\no{\bf Acknowledgements} \break
\no We are grateful to J.~Bagger, S.~Ferrara, D.~Z.~Freedman, J.~Harvey and
E.~Witten for discussions and correspondence, and to E.~Witten for suggesting
this model. This work was supported in part by DOE grant DOE-91ER40618 and NSF
grant PHY89-04035.
\endpage
\refout
\end